\documentclass[letterpaper,english,aps]{revtex4}
\usepackage[T1]{fontenc}
\usepackage[latin1]{inputenc}
\usepackage{amsmath}
\usepackage{graphicx}
\usepackage{amssymb}

\makeatletter
\usepackage{babel}
\makeatother
\begin{document}

\title{
Counting Complex Disordered States by Efficient Pattern Matching: \\
Chromatic Polynomials and Potts Partition Functions
}

\author{Marc Timme$^{1}$, Frank van Bussel$^{1}$, Denny Fliegner$^{1}$,
and Sebastian Stolzenberg$^{2}$}

\affiliation{$^{1}$Max Planck Institute for Dynamics \& Self-Organization, Bunsenstr.~10,
37073 Göttingen, Germany}

\affiliation{$^{2}$Department of Physics, Cornell University, 109 Clark Hall,
Ithaca, New York 14853-2501, USA}

\date{25 November 2008}

\begin{abstract}
Counting problems, determining the number of possible states of a
large system under certain constraints, play an important role in
many areas of science. They naturally arise for complex disordered
systems in physics and chemistry, in mathematical graph theory, and
in computer science. Counting problems, however, are among the hardest
problems to access computationally. Here we suggest a novel method
to access a benchmark counting problem, finding chromatic polynomials
of  graphs. We develop a vertex-oriented symbolic pattern matching
algorithm that exploits the equivalence between the chromatic polynomial
and the zero-temperature partition function of the Potts antiferromagnet
on the same graph. Implementing this bottom-up algorithm using appropriate
computer algebra, the new method outperforms standard top-down methods
by several orders of magnitude, already for moderately sized graphs.
As a first application we compute chromatic polynomials of samples
of the simple cubic lattice, for the first time computationally accessing three-dimensional lattices of physical relevance. The method offers
straightforward generalizations to several other counting problems.
\end{abstract}
\maketitle

Given a set of different colors, in how many ways can one color the
vertices of a graph such that no two adjacent vertices have the same
color? The answer to this question is provided by the chromatic polynomial
of a graph \cite{Birkhoff,Read}, which gives the number of possible
colorings as a function of the number $q$ of colors available. It is a polynomial in $q$ of degree $N$, the number of vertices of the graph. The chromatic polynomial is closely related to other graph invariants e.g. to the reliability and flow polynomials of a network or graph (functions that characterize its communication capabilities) and to the Tutte polynomial. These are of widespread interest in graph theory and computer science and pose similar hard counting problems.

The chromatic polynomial is also of direct relevance to statistical physics as it is equivalent to the zero-temperature partition function of the Potts antiferromagnet \cite{Potts,Wu}: The Potts model \cite{Potts} constitutes a paradigmatic characterization of systems of interacting electromagnetic moments or spins, where each spin can be in one out of $q\geq 2$ states; it thus generalizes the Ising model where $q=2$. For antiferromagnetic interactions, neighboring spins tend to disalign such that at zero temperature, the partition function of the Potts antiferromagnet counts the number of ground states of a spin system just as the chromatic polynomial counts the number of proper colorings of the same graph. For sufficiently large $q$ there are \emph{many} system configurations in which \emph{all} pairwise interaction energies are minimized at zero temperature. Indeed, these systems exhibit a large number of disordered ground states that is exponentially increasing with system size. Thus the Potts model exhibits positive ground state entropy, an exception to the third law of thermodynamics.  Experimentally, complex disordered ground states and related residual entropy at low temperatures have been observed in various systems \cite{Pauling,Parsonage,Ramirez,Broholm,Wills,Fennel}.

Although there are several analytical approaches to find chromatic
polynomials for families of graphs and to bound their values \cite{BiggsMatrixMethod,Birkhoff,Potts,SalasShrock64:011111,SokalAnalytics,SokalAnalytics2001,Wu,Read,Jacobsen,Rocek,SalasSokal},
there is no closed form solution to this counting problem for general
graphs. Algorithmically it is hard to compute the chromatic polynomial,
because the computation time in general increases exponentially with
the number of edges in the graph \cite{HardComputation}. It also
strongly depends on the structure of the graph and rapidly increases
with the graph's size, and the degrees of its vertices, cf. \cite{MezardScience,WeigtMuseum,HartmannBook}.
Therefore, most studies on chromatic polynomials up to date have focused
on small graphs and families of graphs of simple structure and low
vertex degrees, e.g. two-dimensional lattice graphs \cite{SalasSokal,Jacobsen,Rocek}
(an interesting recent attempt to analytically study simple cubic
lattices considered strips with reduced degrees \cite{SalasShrock64:011111}).
In fact, it is not at all straightforward to computationally access
larger graphs with more involved structure, including physically relevant
three-dimensional lattice graphs. Finding the chromatic polynomial
of a graph thus constitutes a challenging, computationally hard problem
of statistical physics, graph theory and computer science (cf. \cite{HardComputation,HartmannBook}).

Below we present a novel, efficient method to compute chromatic polynomials
of larger structured graphs. Representing a chromatic polynomial as
a zero temperature partition function of the Potts antiferromagnet
we transform the computation into a local and vertex-oriented, ordered
pattern matching problem which we then implement using appropriate
computer algebra. In contrast to conventional top-down methods that
represent and process the entire graph (and many modified copies thereof)
from the very beginning, the new method presented here works through
the graph bottom-up and thus processes comparatively small local parts
of the graph only.

Consider a graph $G$ that is defined by a set of $N$ vertices $i\in V=\{1,\ldots,N\}$
and a set of $M:=|E|$ edges $\{ i,j\}\in E$, each edge joining two
vertices $i$ and $j$ which are then called adjacent or neighboring.
This graph is said to be (properly) $q$-colored if every vertex is
given one out of $q$ colors $\{1,2,\ldots,q\}$ such that every two
adjacent vertices have different colors. The number of $q$-colorings
of a graph $G$ is expressed by its chromatic polynomial $P(G,q)$,
a polynomial in $q$ of order $N$ \cite{Read}.  

The deletion-contraction theorem of graph theory \cite{Read} suggests
a simple algorithm to compute the chromatic polynomial of a given
graph recursively. In principle, this algorithm works for arbitrary
graphs and is therefore, with certain improvements, implemented in
general-purpose computer algebra systems such as Mathematica \cite{Mathematica,SkienaOriginal,Pemmaraju}
and Maple \cite{Maple} (cf.\ also \cite{Niejenhuis_et_al,ReadWaterloo}).
 However, applying the theorem recursively the chromatic polynomial
of exponentially many graphs must be found, the (weighted) sum of
which yields the chromatic polynomial of the original graph. This
reflects how hard the problem is algorithmically and severely restricts
the applicability of computational methods, in particular if they
employ standard top-down processing.

\begin{figure}[t]
\begin{centering}\includegraphics[width=70mm,keepaspectratio]{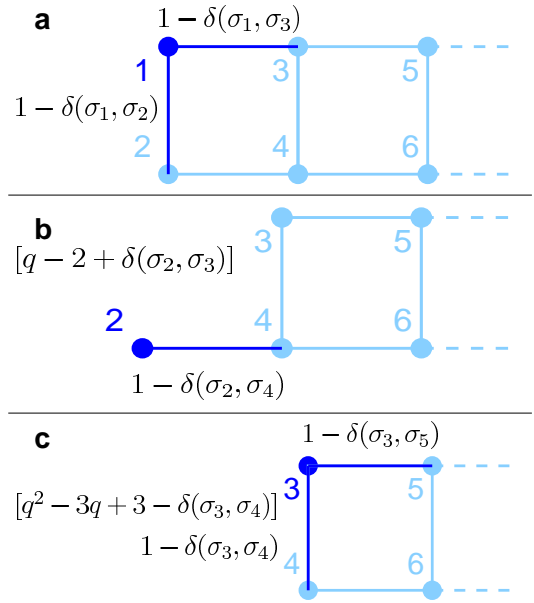}\par\end{centering}

\caption{The bottom-up, local and vertex-oriented nature of the algorithm.
Sequential processing of edges (dark) adjacent to vertices (a) $i=1$,
(b) $i=2$, and (c) $i=3$. The remainder graph (light) is not affected
when processing vertex $i$. Partial partition functions $z_{i-1}$
computed so far (before each vertex step $i$) are shown in square
brackets. \label{fig:algorithm}}
\end{figure}

We now describe our novel algorithm. It is based on the antiferromagnetic
($J<0)$ Potts model \cite{Potts,Wu} with Hamiltonian\begin{equation}
H(\boldsymbol{\sigma})=-J\sum_{\{ i,j\}\in E}\delta_{\sigma_{i}\sigma_{j}}\label{eq:Hamiltonian}\end{equation}
giving the total energy of the system in state $\boldsymbol{\sigma}=(\sigma_{1},\ldots,\sigma_{N})$.
Here individual spins $\sigma_{i}$ can assume $q$ different values
$\sigma_{i}\in\{1,\ldots,q\}$, generalizing he Ising model ($q=2$) \cite{Ising}. Two spins $\sigma_{i}$ and $\sigma_{j}$
on the graph $G$ interact if and only if they are neighboring, $\{ i,j\}\in E$,
and in the same state, $\sigma_{i}=\sigma_{j}$, i.e. the Kronecker-delta
is $\delta_{\sigma_{i}\sigma_{j}}=1$ (otherwise, for any pair $\sigma_{i}\neq\sigma_{j}$,
it is $\delta_{\sigma_{i}\sigma_{j}}=0$). Thus the total interaction energy is minized if all pairs of neighboring spins are in different states.

The partition function $Z(G,q,T)=\sum_{\boldsymbol{\sigma}}\exp(-\beta H(\boldsymbol{\sigma}))$
at positive temperature $T=(k_{\textrm{B}}\beta)^{-1}$, where $k_{\textrm{B}}$
is the Boltzmann constant, can be represented as \begin{equation}
Z(G,q,T)=\sum_{\boldsymbol{\sigma}}\prod_{\{ i,j\}\in E}(1+v\delta_{\sigma_{i}\sigma_{j}})\label{eq:partitionfunction}\end{equation}
where $v=\exp(\beta J)-1\in(-1,0].$ In the  limit $T\rightarrow0$
(implying $\beta J\rightarrow-\infty$ and thus $v\rightarrow-1$)
this  partition function counts the number of ways of arranging the spins $\mathbf{\sigma}$ such that no two adjacent spins are in the same state. Thus the zero temperature partition function (\ref{eq:partitionfunction}) exactly equals \cite{Wu} the chromatic polynomial\begin{equation}
P(G,q)=\lim_{T\rightarrow0}Z(G,q,T)\label{eq:limitT0}\end{equation}
 on the same graph $G$ leading to the representation \cite{Wu,SokalAnalytics}
\begin{equation}
P(G,q)=\sum_{\sigma_{1}=1}^{q}\cdots\sum_{\sigma_{N}=1}^{q}\prod_{\{ i,j\}\in E}(1-\delta_{\sigma_{i}\sigma_{j}})\label{eq:CPsumproduct}\end{equation}
of the chromatic polynomial in terms of sums over products of Kronecker-deltas.

The algorithm exploits this representation by expanding the products
in (\ref{eq:CPsumproduct}) and symbolically evaluating the right
hand side vertex by vertex (cf. Fig. \ref{fig:algorithm}), considering
each individual sum $\sum_{\sigma_{k}}$ as an operator. This operator
interpretation relies on a recently studied algebraic structure of
expressions containing Kronecker-deltas \cite{Kozen}. Here such an
operator has the simple actions \begin{equation}
\sum_{\sigma_{k}}1=q,\label{eq:sum0}\end{equation}
\begin{equation}
\sum_{\sigma_{k}}\delta_{\sigma_{k}\sigma_{j_{1}}}=1,\label{eq:sum1}\end{equation}
 \begin{equation}
\sum_{\sigma_{k}}\delta_{\sigma_{k}\sigma_{j_{1}}}\delta_{\sigma_{k}\sigma_{j_{2}}}=\delta_{\sigma_{j_{1}}\sigma_{j_{2}}}\,,\label{eq:sum2}\end{equation}
and for an arbitrary number $r\in\mathbb{N}_{0}$ of factors, \begin{equation}
\sum_{\sigma_{k}}\delta_{\sigma_{k}\sigma_{j_{1}}}\cdots\delta_{\sigma_{k}\sigma_{j_{r}}}=\delta_{\sigma_{j_{1}}\sigma_{j_{2}}}\cdots\delta_{\sigma_{j_{r-1}}\sigma_{j_{r}}}\label{eq:sum:delta:r}\end{equation}
if the $j_{\rho}$, $\rho\in\{1,\ldots,r\}$, are pairwise distinct
and all $j_{\rho}\neq k$.

For illustration consider the chromatic polynomial\begin{equation}
P(G,q)=\sum_{\sigma_{3}=1}^{q}\sum_{\sigma_{2}=1}^{q}\sum_{\sigma_{1}=1}^{q}(1-\delta_{\sigma_{1}\sigma_{2}})(1-\delta_{\sigma_{1}\sigma_{3}})(1-\delta_{\sigma_{2}\sigma_{3}})\label{eq:CP:triangle}\end{equation}
 of a triangular (complete) graph comprised of $N=3$ vertices and
$M=3$ edges. We start at vertex $i=1$ by expanding the relevant
product \begin{align}
p_{1} & =(1-\delta_{\sigma_{1}\sigma_{2}})(1-\delta_{\sigma_{1}\sigma_{3}})\label{eq:p1}\\
 & =1-\delta_{\sigma_{1}\sigma_{2}}-\delta_{\sigma_{1}\sigma_{3}}+\delta_{\sigma_{1}\sigma_{2}}\delta_{\sigma_{1}\sigma_{3}}\label{eq:p2}\end{align}
that is comprised of all factors that contain $\sigma_{1}$. (We note
that already Birkhoff \cite{Birkhoff} in 1912 used closely related
expansions to theoretically derive an alternative representation of
chromatic polynomials.) Symbolically applying the above replacement
rules (\ref{eq:sum:delta:r}) yields a {}``partial partition function''
\begin{eqnarray}
z_{1} & = & \sum_{\sigma_{1}=1}^{q}p_{1}\label{eq:z11}\\
 & = & q-2+\delta_{\sigma_{2}\sigma_{3}}\,,\label{eq:z13}\end{eqnarray}
and thus $P(G,q)=\sum_{\sigma_{3}=1}^{q}\sum_{\sigma_{2}=1}^{q}z_{1}(1-\delta_{\sigma_{2}\sigma_{3}})$.
Proceeding with the vertices $i=2$ and $i=3$ in a similar fashion,
we obtain $p_{2}=(q-2+\delta_{\sigma_{2}\sigma_{3}})(1-\delta_{\sigma_{2}\sigma_{3}})$,
$z_{2}=\sum_{\sigma_{2}=1}^{q}p_{2}=(q-1)(q-2)$, $p_{3}=z_{2}$,
and reduce the chromatic polynomial to the final result $P(G,q)=z_{3}=\sum_{\sigma_{3}=1}^{q}p_{3}=q(q-1)(q-2)$,
successively. 

For a general graph $G$ on $N$ vertices, the algorithm is analogous
to the example. First define $z_{0}=1$. Then, passing through the
vertices from $i=1$ sequentially up to $i=N$,

\begin{enumerate}
\item construct and expand $p_{i}=z_{i-1}\prod_{\{ i,j\}\in E}(1-\delta_{\sigma_{i}\sigma_{j}})$
where the product is over all edges incident to $i$ that have not
been considered before, i.e.\ $j>i$;
\item \emph{symbolically} evaluate the sum $z_{i}=\sum_{\sigma_{i}}p_{i}$
applying the simple rules (\ref{eq:sum0})-(\ref{eq:sum:delta:r})
given above. 
\end{enumerate}
These operations are local and vertex oriented in the sense that they
jointly consider all edges $\{ i,j\}$ incident to an individual vertex
$i$ at any one time. A major advantage of this bottom-up algorithm
is that all edges that are not currently processed are kept outside
the computations until they are needed, quite in contrast to standard
top-down deletion-contraction algorithms. If a graph $G$ has a layered
structure,\begin{equation}
G=\bigcup_{\nu}H_{\nu}\label{eq:layeredG}\end{equation}
with layers $H_{\nu}$, constituting samples of periodic lattices
or aperiodic graphs, the vertices $i$ are selected (i.e.\ numbered)
layer by layer such that the operations only affect a particularly
small portion of the graph at once (Fig.~\ref{fig:algorithm}). These
graphs have bounded tree widths, cf. \cite{Andrzejak,Noble}.  More
generally, vertices are numbered appropriately beforehand, for instance,
using minimal band width of the graph as a heuristic criterion \cite{WestBandWidth}.

The computation of the chromatic polynomial has been reduced to a
process of alternating expansion of expressions and symbolically replacing
terms in an appropriate order. In the language of computer science,
these operations are represented as the expanding, matching, and sorting
of patterns, making the algorithm suitable for computer algebra programs
optimized for pattern matching.

To fully exploit the capabilities of this algorithm, we implemented
it using the language Form \cite{Form,Vermaseren} which is specialized
to large scale symbolic manipulation problems and as such a successful
standard tool for, e.g., Feynman diagram evaluation in precision high-energy
physics \cite{Laporta,Misiak}.

A practically relevant measure for the speed of our method is the
total CPU time \textbf{}$t$  it needs for a specific calculation.
For hard counting problem, one generally expects an exponential increase
$t\approx A\exp(\alpha m)\,$ with the size $m\gg1$ of the problem,
here defined as the number $m=M$ of edges for chromatic polynomials
of general graphs. For graphs with bounded tree width \cite{Andrzejak,Noble}
the solution time of the counting problem typically only grows exponentially
with the width of the graph, i.e. in the square of the number of vertices
in the subgraphs $H_{\nu}$. The factor $\alpha$ in the exponent
determines the scaling of the computational time with problem size
and measures the efficiency of the algorithm, whereas the prefactor
$A$ fixes the absolute time needed and depends, among others, on
the software environment and hardware used.

\begin{figure}[t]
\begin{centering}\includegraphics[width=80mm,keepaspectratio]{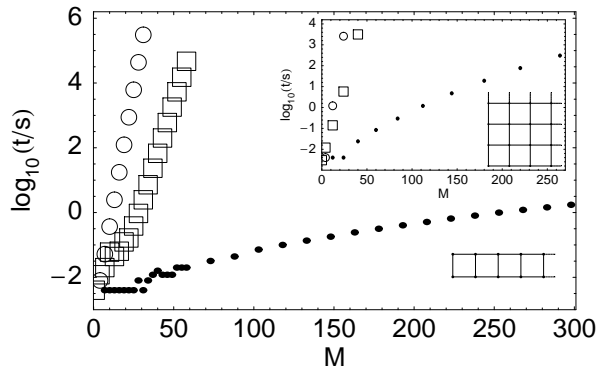}\par\end{centering}

\caption{Computation time $t$ (in ms) for square lattice samples of sizes
$2\times n$ (main panel) and $n\times n$ (inset) increases with
the number of edges $M$ of the graph. The new method ($\bullet$)
drastically outperforms standard methods used in Mathematica ($\bigcirc$)
and Maple \textbf{}($\square$), both with respect to the scaling
of the algorithm (quantified by the local slope) and the absolute
effectiveness (quantified by the absolute times needed), even for
moderately sized graphs. \label{fig:computationtimeREGULAR}}
\end{figure}

\begin{figure}[t]
\begin{centering}\includegraphics[width=140mm,keepaspectratio]{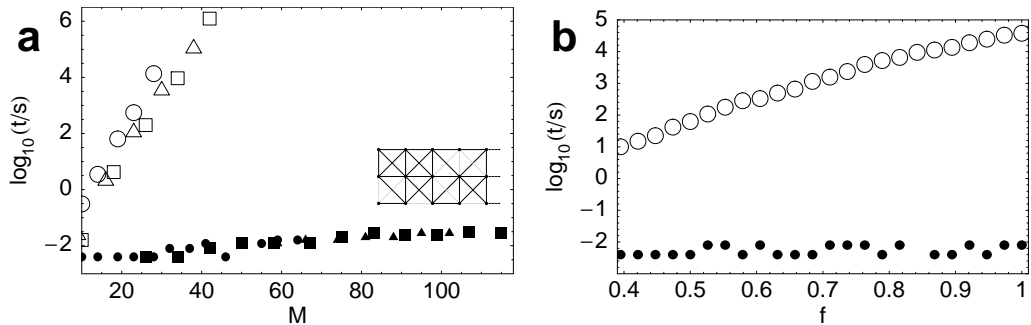}\par\end{centering}

\caption{Computation time $t$ (in ms) for randomly edge-diluted square lattice
samples with next nearest neighbor (nnn) interactions (displayed as
a cartoon inset in panel (a)). The new method (filled symbols) strongly
outperforms standard deletion-contraction method \textsc{(}Mathematica,
open symbols). (a) Computation time vs. the actual total number $M$
of edges for samples of $3\times n$ vertices where each original
edge has been deleted independently with probability $p=0.1$ (circles)
$p=0.2$ (triangles) and $p=0.5$ (squares). (b) Computation time
vs. the fraction $f$ of edges present. Edges are sequentially randomly
removed independently, starting from an undiluted $3\times5$ square
lattice graph ($f=1)$ with nnn interactions; the graph is diluted
until before it becomes disconnected below $f\approx0.4$. Whereas
the computation times using the new method are still at fluctuation
level ($t<10^{-2}s)$, standard methods take at factor of $10^{3}$
to $10^{7}$ longer. \label{fig:computationtimeDILUTED}}
\end{figure}

To compare our method to existing ones, we first computed chromatic
polynomials of samples of the two-dimensional square lattice with
free boundary conditions ($2\times n$ strips that have $M=3n-2$
edges and $n\times n$ patches that have $M=2n(n-1)$ edges). The
total computation times $t$ have been measured as a function of $M$
for the new method as well as for the standard methods used in Mathematica
\cite{Mathematica,SkienaOriginal} and Maple \cite{Maple}, respectively.
Figure \ref{fig:computationtimeDILUTED} shows that the scaling $\alpha$
of the algorithm (given by the local slope of the data points in the
logarithmic plot) of our new method is markedly better than the one
found for the standard deletion-contraction methods. This implies
that the new method outperforms these standard computational methods
in the absolute computation time by several orders of magnitude already
for moderately sized graphs (e.g. about six orders of magnitude for
$n=10$ i.e. $M=28)$. With increasing graph size, the advantages
of our method become more pronounced. For example, for the $2\times100$
strip of the square lattice ($M=298$ edges) the pattern matching
method needs a computation time of the order of $t\approx2s$ whereas
extrapolation of the data shown in Fig.~\ref{fig:computationtimeDILUTED}
indicates that the same problem is not computationally accessible
using the standard deletion-contraction methods implemented in Mathematica.
Second, in contrast to transfer matrix or other analytical recurrence
methods \cite{SokalAnalytics,SokalAnalytics2001,SalasSokal}, the
above method also works in a simple way for graphs with non-identical
subgraphs $H_{\nu}$, such as randomly diluted lattices. The same
comparison for randomly diluted $3\times n$ square lattice samples
with next-nearest neighbor interactions (Fig. \ref{fig:computationtimeDILUTED})
confirms the pronounced outperformance and moreover illustrates the
general applicability of our method, also compared to recursive analytical
methods. 

\begin{figure}[tp]
\begin{centering}\includegraphics[width=100mm,keepaspectratio]{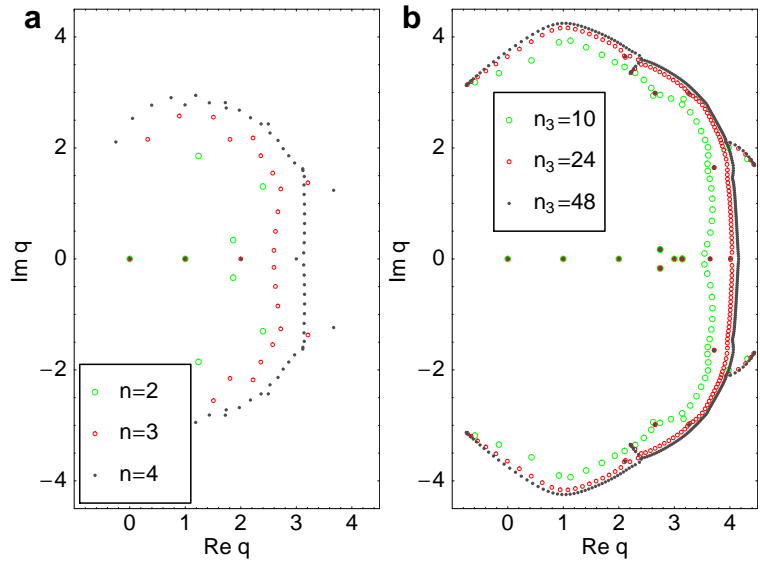}\par\end{centering}

\caption{Complex zeroes representing chromatic polynomials of samples of the
simple cubic lattice (a) $n\times n\times n$ Cartesian samples with
free boundary conditions,  and (b)  $2\times4\times n_{3}$ diagonal
samples with periodic transverse boundary conditions, with up to $M=1128$
edges. \label{fig:chromaticzeroes}}
\end{figure}

As a first application to an open hard counting problem, we now turn
to three-dimensional lattices of direct physical relevance. First,
we consider $n\times n\times n$ samples of the simple cubic lattice
with free boundary conditions, which have $N=n^{3}$ vertices and
$M=3n^{2}(n-1)$ edges.  We found chromatic polynomials up to $n=4$
($N=64$, $M=144$). A representation of the chromatic polynomial
$P(G,q)$ in terms of its $N$ complex zeroes $q_{1},\ldots,q_{N}$
is shown in Fig.~\ref{fig:chromaticzeroes}a for $n=3$ and $n=4$.
We further consider simple cubic lattice strips that extend in the
\emph{diagonal (111)} direction with periodic boundaries in the two
other (transverse) directions.  This keeps the number of vertices
within one layer low at the same time allowing for a large number
$N_{c}$ of vertices with the same degree (equal to six) as vertices
in the infinite lattice, a fact that is heuristically known to be
essential for a rapid convergence towards the thermodynamic limit
($N,M\rightarrow\infty$).  The largest three-dimensional sample
graphs shown in Figure \ref{fig:chromaticzeroes}b  have $N=384$
vertices, $M=1128$ edges and a fraction $N_{c}/N=368/384\approx0.96$
of vertices with correct degree, (as compared to $N_{c}/N=8/64\approx0.13$
for the $4\times4\times4$ sample extending along the Cartesian axes
and to $N_{c}=0$ for previous attempts to address three-dimensional
lattices). The computation time was approximately 11 hours on a single
Linux machine with an Intel Pentium 4, 2.8 GHz-32 bit processor. 

In summary we have presented a novel method to calculate chromatic
polynomials of graphs. Using the partition function representation,
it proceeds vertex by vertex employing an \textit{a priori} reduction
to local operations only and is thus particularly suited for graphs
exhibiting a layered structure. The method combines a symbolic bottom-up
algorithm, which is based on systematic term-wise expansion and pattern
matching, with an appropriate computer algebra program \cite{Form,Vermaseren}.
Our method is applicable to general types of graphs, including graphs
with bounded and unbounded tree widths as well as randomized graphs.
We demonstrated by several sets of examples that it drastically outperforms
existing standard methods for all these types of graphs. As a practical
application, we computed chromatic polynomials for samples of the
simple cubic lattice, for the first time computationally accessing
three-dimensional lattices of physical relevance.

Since the main ideas underlying our method are simple to apply, they
may be generalized in a straightforward way and also be transferred
to other challenging counting problems. Among others, one may compute
quantitative measures relevant in computer science that give information
about the communication capabilities of a network, such as (i) the
flow polynomial and (ii) the reliability polynomial \cite{FlowPolynomials,ReliabilityPolynomials}.
It is equally possible to determine (iii) ferromagnetic and (iv) positive
temperature partition functions of statistical physics \cite{ChangposT,Haeggkvist,Chen}
and, equivalently, (v) the Tutte polynomial (of two variables $q$
and $v$, cf.\ Eq.~\ref{eq:partitionfunction}), valuations of which
directly result in the number of spanning subgraphs, the number of
spanning trees, and other invariants of a graph \cite{SokalAnalytics,SokalAnalytics2001}.
Of course, applications in graph theory may include studies  of families
of graphs where computational results seemed impossible so far, because
the computational effort is substantially reduced. As the new method
yields exact results not only for the final solution (in our examples,
the chromatic polynomial) but also in the intermediate steps (the
partial partition functions above), it may moreover be combined with
analytical tools \cite{ChangposT,Chen,ReliabilityPolynomials,SalasShrock64:011111,SalasSokal,Jacobsen,Rocek}
to obtain unprecedented results for various classes of graphs. Finally,
the method can easily be implemented in parallel computations. Taken
together, the novel bottom-up pattern matching algorithm combined
with specialized computer algebra presented here constitutes a promising
starting point to access a number of challenging, computationally
hard counting problems from statistical physics, graph theory and
computer science.

\begin{acknowledgments}
M.T. thanks R. Shrock for introducing him to the subject. We thank
C. Ehrlich, T. Geisel, L. Hufnagel, D. Kozen, H. Schanz, R. Shrock,
A. Sokal, J. Vermaseren and M. Weigt for helpful comments. This work
was supported by the Max Planck Society via a grant to M.T. and the
Max Planck Advisory Committee for Electronic Data Processing (BAR).

\end{acknowledgments}


\begin{thebibliography}{22}
\bibitem{Birkhoff}Birkhoff, G. D. A Determinant Formula for the Number
of Ways of Coloring a Map. \emph{Ann. of Math.} 14, 42 (1912).

\bibitem{Read}Read, R.C. \emph{Chromatic Polynomials}, in \emph{Selected
Topics in Graph Theory 3}, edited by L.W. Beineke and R.J. Wilson,
Academic Press, London (1988).

\bibitem{Potts}Potts, R. B. Some Generalized Order-Disorder Transformations.
\emph{Proc. Camb. Phil. Soc.} \textbf{48}, 106 (1952); 

\bibitem{Wu}Wu, F. Y. The Potts model. \emph{Rev. Mod. Phys.} \textbf{54},
235 (1982); Erratum: The Potts model. \textbf{55}, 315(E) (1983).

\bibitem{Pauling}Pauling, L. \emph{The Nature of the Chemical Bond}
(Cornell University Press, Ithaca, New York, 1960).

\bibitem{Parsonage}Parsonage, N. G. \& Staveley, L. A. K. \emph{Disorder
in Crystals} (Oxford University Press, Oxford, England, 1978).

\bibitem{Ramirez}Ramirez, A. P. , Espinosa, G. P., \& Cooper, A.
S. Strong frustration and dilution-enhanced order in a quasi-2D spin
glass. \emph{Phys. Rev. Lett.} \textbf{64}, 2070 (1990).

\bibitem{Broholm}Broholm, C., Aeppli, G., Espinosa, G. P., \& Cooper,
A. S. Antiferromagnetic Fluctuations and Short-Range Order in a Kagomé
Lattice. \emph{Phys. Rev. Lett.} \textbf{65}, 3173 (1990).

\bibitem{Wills}Wills, A. S., Harrison, A., Mentink, S. A. M., Mason,
T. E., \& Tun, Z. Magnetic correlations in deuteronium jarosite, a
model $S=5/2$ Kagomé antiferromagnet. \emph{Europhys. Lett.} \textbf{42},
325 (1998).

\bibitem{Fennel}Fennell, T., Bramwell, S.T., McMorrow, D.F., Manuel,
P., \& Wildes, A.R. Pinch points and Kasteleyn transitions in kagome
ice. \emph{Nature Phys.} \textbf{3}, 566 (2007).

\bibitem{SalasShrock64:011111}Salas, J. \& Shrock, R. Exact $T=0$
partition functions for Potts antiferromagnets on sections of the
simple cubic lattice. \emph{Phys. Rev. E} \textbf{64}, 011111 (2001).

\bibitem{SokalAnalytics}Sokal, A. D. Chromatic polynomials, Potts
models and all that. \emph{Physica A} \textbf{279}, 324 (2000).

\bibitem{SokalAnalytics2001}Sokal, A. D. Bounds on the Complex Zeros
of (Di)Chromatic Polynomials and Potts-Model Partition Functions.
\emph{Combin. Probab. Comput.} \textbf{10}, 41 (2001).

\bibitem{BiggsMatrixMethod}Biggs, N. A Matrix Method for Chromatic
Polynomials. \emph{J. Combin. Theory, Series B.} \textbf{82}, 19
(2001).

\bibitem{SalasSokal}Salas, J. \& Sokal, A. D. Transfer Matrices and
Partition-Function Zeros for Antiferromagnetic Potts Models. I. General
Theory and Square-Lattice Chromatic Polynomials. \emph{J. Stat. Phys.}
\textbf{104}, 609 (2001).

\bibitem{Jacobsen}Jacobsen, J. L. \& Salas, J. Transfer Matrices
and Partition-Function Zeros for Antiferromagnetic Potts Models. II.
Extended Results for Square-Lattice Chromatic Polynomial. \emph{J.
Stat. Phys.} \textbf{104}, 701 (2001).

\bibitem{Rocek}Rocek, M., Shrock, R., \& Tsai, S.-H. Chromatic polynomials
for families of strip graphs and their asymptotic limits. \emph{Physica
A} \textbf{252}, 505 (1998).

\bibitem{HardComputation}Papadimitriou, C. \emph{Computational Complexity,}
Addison-Wesley (1994).

\bibitem{HartmannBook}Hartmann, A. K. \& Weigt, M. \emph{Phase Transitions
in Combinatorial Optimization Problems,} Wiley-VCH (2005).

\bibitem{MezardScience}Mezard, M., Parisi, G., \& Zecchina, R. Analytic
and Algorithmic Solution of Random Satisfiability Problems. \emph{Science}
\textbf{297}, 812 (2002).

\bibitem{WeigtMuseum}Weigt, M. \& Hartmann, A. K. Number of Guards
Needed by a Museum: A Phase Transition in Vertex Covering of Random
Graphs. \emph{Phys. Rev. Lett.} \textbf{84}, 6118 - 6121 (2000).

\bibitem{Mathematica}\textsc{Mathematica,} Release 6, Wolfram Research
Inc. (2007).

\bibitem{SkienaOriginal}Skiena, S.S. \emph{Implementing Discrete
Mathematics}, Addison-Wesley (1990).

\bibitem{Pemmaraju}Pemmaraju, S.V. \& Skiena, S.S. \emph{Computational
Discrete Mathematics}. Perseus Books (2003).

\bibitem{Maple}\textsc{Maple,} Release 9, Waterloo Maple Inc. (2003).

\bibitem{Niejenhuis_et_al}Nijenhuis, A. \& Wilf, H. S. \textit{Combinatorial
Analysis}, Academic Press (1975).

\bibitem{ReadWaterloo}Read, R. C. \textit{An improved method for
computing chromatic polynomials of sparse graphs}, Research Report
CORR 87-20, Univ. of Waterloo (1987).

\bibitem{Ising} Ising E., Z. Physik (1925).

\bibitem{Kozen}Kozen, D. \& Timme, M. Indefinite Summation and the
Kronecker Delta. Technical Report, Cornell University, http://hdl.handle.net/1813/8352
(2008).

\bibitem{Noble}Noble, S. D. Evaluating the Tutte Polynomial for Graphs
of Bounded Tree-Width.  \emph{Combin. Probab. Comput.} \textbf{7},
307 (1998).

\bibitem{Andrzejak}Andrzejak, A. An algorithm for the Tutte polynomials
of graphs of bounded treewidth.  \emph{Discrete Math.}, \textbf{190},
39 (1998).

\bibitem{WestBandWidth}West, D. B. \emph{Introduction to Graph Theory.}
Prentice Hall (2000). 

\bibitem{Form}\textsc{Form}, Version 3.1, available free of charge
from http://www.nikhef.nl/\textasciitilde{}form (2002).

\bibitem{Vermaseren}Vermaseren, J. New features of FORM. http://arxiv.org:
math-ph/0010025 (2000).

\bibitem{Laporta}Laporta, S. \& Remiddi, E. The analytical value
of the electron $(g-2)$ at order $\alpha^{3}$ in QED. \emph{Phys.
Lett. B} \textbf{379}, 283 (1996).

\bibitem{Misiak}Misiak, M. \& Steinhauser, M. Three-loop matching
of the dipole operators for $b\rightarrow s\gamma$ and $b\rightarrow sg$.
\emph{Nucl. Phys. B} \textbf{683}, 277-305 (2004).

\bibitem{MPSolve}Bini, D. A. \& Fiorentino, G. \emph{Numer. Algorithms}
\textbf{23}, 127 (2000); \textit{Numerical Computation of Polynomial
Roots:} \textsc{MPSolve} \textit{2.0}, Frisco report of the Numerical
Algorithms Group Ltd. (1998), available via http://www.nag.co.uk/projects/frisco/frisco/node8.htm.

\bibitem{ChangposT}Chang, S.-C. \& Shrock, R. Exact $T=0$ partition
functions for Potts antiferromagnets on sections of the simple cubic
lattice. \emph{Phys. Rev. E} \textbf{64}, 066116 (2001).

\bibitem{Haeggkvist}Häggkvist, R. \emph{et al.} Computation of the
Ising partition function for two-dimensional square grids\emph{,}
\emph{Phys. Rev. E} \textbf{69}, 046104 (2004).

\bibitem{Chen}Chen, C.-N., Hu, C.-K., \& Wu, F. Y. Partition Function
Zeros of the Square Lattice Potts Model. \emph{Phys. Rev. Lett.} \textbf{76},
169 (1996).

\bibitem{FlowPolynomials}Jackson, B. Zeros of chromatic and flow
polynomials of graphs.  \emph{J. Geom.} \textbf{76}, 95 (2003).

\bibitem{ReliabilityPolynomials}Chang, S.-C. \& Shrock, R. Reliability
Polynomials and Their Asymptotic Limits for Families of Graphs. 
\emph{J. Stat. Phys.} \textbf{112}, 1019 (2003).
\end{thebibliography}
\end{document}